\documentclass[%showpacs, showkeys,
12pt,
preprint,preprintnumbers,nofootinbib,
groupedaddress,superscriptaddress,amsmath,amssymb]{revtex4}
\usepackage{graphicx}% Include figure files
\usepackage{dcolumn}% Align table columns on decimal point
\usepackage{bm}% bold math
\usepackage{amssymb}
\usepackage{amsmath}
\usepackage{epsfig}    
\usepackage{color}
\usepackage{slashed}
\usepackage{hhline}
\def\be{\begin{equation}}
\def\ee{\end{equation}}
\newcommand{\bea}{\begin{eqnarray}}
\newcommand{\eea}{\end{eqnarray}}
\newcommand{\nn}{\nonumber}

\numberwithin{equation}{section}

\begin{document}

{\begin{flushright}{KIAS-P17120}
\end{flushright}}

%%%%%%%%%
\title{A radiative seesaw model with higher order terms \\
 under an alternative $U(1)_{B-L}$}
%\preprint{KIAS-P14078}
%

\author{Takaaki Nomura}
\email{nomura@kias.re.kr}
\affiliation{School of Physics, KIAS, Seoul 02455, Korea}

\author{Hiroshi Okada}
\email{macokada3hiroshi@cts.nthu.edu.tw}
\affiliation{Physics Division, National Center for Theoretical Sciences, Hsinchu, Taiwan 300}

\date{\today}

\begin{abstract}
We propose a model based on an alternative $U(1)_{B-L}$ gauge symmetry with 5 dimensional operators in the Lagrangian, and we construct the neutrino masses at one-loop level, and discuss lepton flavor violations, dark matter, and the effective number of neutrino species due to two massless particles in our model. Then we search allowed region to satisfy the current experimental data of neutrino oscillation and lepton flavor violations without conflict of several constraints such as stability of dark matter and the effective number of neutrino species, depending on normal hierarchy and inverted one.
\end{abstract}
\maketitle
\newpage

\section{Introduction}
%%%
Unlikely to the gauged $U(1)_{B-L}$ models inspired by grand unified theories such as $SO(10)$~\cite{Babu:1992ia} and $E_6$~\cite{Mohapatra:1986bd},
alternative gauged $U(1)_{B-L}$ models with $(-4,-4,5)$ charges for three right-handed neutrinos seem not to be embedded in any larger groups~\cite{Montero:2007cd}.~\footnote{Note here that the same charge assignments for the right-handed neutrinos are applied to the different neutrino mass mechanisms of Dirac neutrino or inverse seesaw model~\cite{Ma:2014qra, Ma:2015raa}, introducing $S_3$ flavor symmetry. }
Nevertheless, this kind of models also possess a vast of unique potential to extend the standard model (SM) in aspects of neutrino sector, phenomenologies of massless bosons(fermions), dark matter (DM) sector, leptogenesis, collider physics at Large Hadron Collider, and their related issues~\cite{Patra:2016ofq, Singirala:2017see, Nomura:2017vzp, Nomura:2017jxb, Nomura:2017wxf, Nanda:2017bmi, Singirala:2017cch, Geng:2017foe}.

In the alternative gauged $U(1)_{B-L}$ model, we do not have Yukawa interactions among the SM Higgs, the SM lepton doublets and right-handed neutrinos due to the charge assignments. Thus neutrino mass generation is not trivial compared to original $U(1)_{B-L}$ models. Also five dimensional Weinberg operator for active neutrino mass is not allowed by the gauge symmetry.
It is then inevitable to investigate some mechanisms of generating neutrino masses, for examples, considering
effective operators at higher order and/or radiative seesaw model at loop level.
%%%
It is Ma model~\cite{Ma:2006km} that is one of the minimal realizations radiatively to induce the neutrino masses including DM.
One of the advantages is to make the hierarchy of related dimensionless couplings milder than the tree-level neutrino masses. 
However, we still need a rather small coupling constant ($\lambda_5\sim10^{-4}$) associated with a quartic interaction between the SM Higgs and a inert doublet $\eta$, $(H^\dagger \eta)^2$, in scalar potential, if once Yukawa couplings are taken to be $\mathcal{O}(1)$ scale. 
%where $\lambda_5$ corresponds to $\lambda_{H\eta}''$ in our model that will be explained in the next section.
To obtain Yukawa couplings not much smaller than $\mathcal{O}(1)$, one might achieve the way of introducing a higher order term which provides $\lambda_5$ coupling or introducing a concrete structure to generate $\lambda_5$ coupling at loop level. The latter case could be achieved by introducing more symmetries with new fields that mediate inside a loop diagram for generating $\lambda_5$. As example of the successful model, see ref.~\cite{Aoki:2013gzs}.  In our analysis we introduce higher order terms of non-renormalizable level which are invariant under our alternative $U(1)_{B-L}$ gauge symmetry, assuming these terms come from effects at higher scale characterized by $\Lambda$. In principle our effective operators could be realized by generating at loop level with some heavy particle contents. 
%%%

%%%%%% several interesting and original features are possessed in 
In this letter, we consider an invariant Lagrangian up to five dimensional effective terms under the alternative gauged $U(1)_{B-L}$
in which neutrino mass matrix is induced at one-loop level~\cite{Ma:2006km}. 
%%%
 Also we achieve $\lambda_5$ as a result of five dimensional operator (instead of two loop neutrino mass models),
which plays an role in relaxing scale hierarchy of our parameters by suppression due to the cut-off mass scale, although 
the cut-off scale is arbitrary. This is also one of the promising ideas to obtain similar order of parameters~\cite{Okada:2012np, Kajiyama:2012xg}, as discussed above.
Then we formulate the lepton flavor violations (LFVs), boson and fermion sector.
%%%
In addition, we discuss the possibility of DM and the effective number of neutrino species, since we have two massless physical particles; goldstone boson(GB) and neutral fermion.
The GB is a consequence of two charge differences among three right-handed neutrinos, and  the massless neutral fermion (that is identified as the lightest one of three right handed neutrinos) originates from our specific texture of mass matrix for the right handed neutrinos.
With such massless particles, one has to investigate more concretely whether the lifetime of DM is enough long compared to the age of Universe and these massless particles affect the effective number of neutrino species or not.

This paper is organized as follows.
In Sec.~II, we show our model, 
and formulate the scalar sector, $Z'$ boson, exotic neutral fermion, dark matter sector, and the effective number of neutrino species.
%neutral fermion sector, boson sector, lepton sector, and dark matter sector. Also we analyze the relic density of DM without conflict of direct detection searches, and
Then we carry out global analysis. Finally we conclude and discuss in Sec.~III.
%\newpage

%%%%%%%%%%%%%%%%%%%%%%%%%%%%%%%%%%%%%
%\section{The Model}
%\subsection{Model setup}

 \begin{widetext}
\begin{center} 
\begin{table}[b]%[tbc]
%\begin{tiny}
\begin{tabular}{|c||c|c|c|c|c||c|c|}\hline\hline  
%&\multicolumn{5}{c||}{SM leptons} & \multicolumn{3}{c|}{Exotic fermions} \\\hline
Fermions& ~$Q_L$~ & ~$u_R$~ & ~$d_R$~ &~$L_L$~ & ~$e_R$~ & ~$N_{R_i}$~ & ~$N_{R_3}$~ 
\\\hline 
$SU(3)_C$ & $\bm{3}$  & $\bm{3}$  & $\bm{3}$  & $\bm{1}$  & $\bm{1}$  & $\bm{1}$  & $\bm{1}$  \\\hline 
 %%%
 $SU(2)_L$ & $\bm{2}$  & $\bm{1}$  & $\bm{1}$ & $\bm{2}$ & $\bm{1}$  & $\bm{1}$ & $\bm{1}$   \\\hline 
 %%%
$U(1)_Y$ & $\frac16$ & $\frac23$  & $-\frac{1}{3}$ & $-\frac12$  & $-1$ & $0$ & $0$    \\\hline
 %%%
 $U(1)_{B-L}$ & $\frac13$ & $\frac13$  & $\frac13$ & $-1$  & $-1$   & $-4$   & $5$   \\\hline
  %%%
 $Z_{2}$ & $+$ & $+$  & $+$ & $+$  & $+$   & $-$   & $-$   \\\hline
 %%%
\end{tabular}
%%%%%%%%%%%
%\begin{tabular}{|c||c|c|c|c|c|}\hline\hline
%  Bosons  &~ $H$  &~ $\eta$  ~ &~ $s$~ &~ $\varphi_1$ &~ $\varphi_2$ \\\hline
%$SU(3)_C$ & $\bm{1}$  & $\bm{1}$  & $\bm{1}$  & $\bm{1}$ & $\bm{1}$ \\\hline 
%$SU(2)_L$ & $\bm{2}$ & $\bm{2}$  & $\bm{1}$ & $\bm{1}$ & $\bm{1}$  \\\hline 
%$U(1)_Y$ & $\frac12$ & $\frac12$  & $0$ & $0$ & $0$    \\\hline
% $U(1)_{B-L}$ & $0$ & $-3$ & $4$ & $1$  & $8$  \\\hline
%\end{tabular}%
%%%%%%%%%%%
\caption{Field contents of fermions and bosons and their charge assignments under $SU(3)_C\times SU(2)_L\times U(1)_Y\times U(1)_{B-L} \times Z_2$, where $i=1,2$.}\label{tab:1}
% \end{tiny}
\end{table}\end{center}\end{widetext}

%\if0
\begin{table}[b]
\centering {\fontsize{10}{12}
\begin{tabular}{|c||c|c|c|c|}\hline\hline
  Bosons  &~ $H$~  & ~$\eta$~ & ~$\varphi_5$~ & ~$\varphi_6$~ \\\hline
$SU(3)_C$ & $\bm{1}$  & $\bm{1}$   & $\bm{1}$ & $\bm{1}$ \\\hline 
$SU(2)_L$ & $\bm{2}$ & $\bm{2}$  & $\bm{1}$ & $\bm{1}$  \\\hline 
$U(1)_Y$ & $\frac12$ & $\frac12$ & $0$ & $0$    \\\hline
 $U(1)_{B-L}$ & $0$ & $-3$ & $5$  & $6$  \\\hline
  $Z_{2}$ & $+$ & $-$ & $+$  & $+$  \\\hline
\end{tabular}%
} 
\caption{Field contents of bosons and their charge assignments under $SU(3)_C\times SU(2)_L\times U(1)_Y\times U(1)_{B-L} \times Z_2$, where $\eta$ does not have VEV.}
\label{tab:2}
\end{table}
%\fi

\section{ Model setup and phenomenologies}
In this section, we introduce our model.
First of all, we impose an additional $U(1)_{B-L} \times Z_2$ gauge symmetry with  three right-handed neutral fermions $(N_{R_1},N_{R_2},N_{R_3})$
where the right-handed neutrinos have $U(1)_{B-L}$ charge $-4$, $-4$ and $5$. Then all the anomalies we have to consider are  $U(1)_{B-L}^3$, and $U(1)_{B-L}$, which are found to be zero~\cite{Singirala:2017see}.
On the other hand, even when we introduce two types of isospin singlet bosons $\varphi_5$ and $\varphi_6$ in order to acquire nonzero Majorana masses after the spontaneous symmetry breaking of  $U(1)_{B-L}$, one cannot find active neutrino masses due to the absence of Yukawa term $\bar L_L \tilde H N_R$.
Thus we introduce an isospin doublet inert boson $\eta$ with nonzero $U(1)_{B-L}$ charge which does not develop vacuum expectation value (VEV), and neutrino masses are induced at one-loop level as shown in Fig.~\ref{fig:diagram}. 
%%%
Here $Z_2$ symmetry plays a role in forbidding 5-dimensional Yukawa terms; $\bar L_{L_a} \tilde H N_{R_{1,2}}\varphi^*_5$ and $\bar L_{L_a} \tilde H N_{R_3}\varphi^*_6$ 
 since we require neutrino mass matrix is generated at only one-loop level with effects of higher dimensional operators.
%%%
Field contents and their assignments for fermions and scalar fields are respectively given by Table~\ref{tab:1} and \ref{tab:2}.
% Notice here that all the anomalies related to $U(1)_Y$; $U(1)_Y^2\times U(1)_{B-L}$ and  $U(1)_Y\times U(1)_{B-L}^2$, are zero. 
%\subsection{Yukawa interactions and scalar sector}{\it Yukawa Lagrangian}:
Under these symmetries, the Lagrangian including five dimensional effective terms for lepton sector and Higgs potential are respectively given by 
\begin{align}
-{\cal L}_{L}&=
%\nn\\&
\label{eq:Yukawa}
y_{\ell_{a}}\bar L_{L_a} e_{R_a} H + y_{\nu_{ai}} \bar L_{L_a} \tilde\eta N_{R_i}
%+y_{N_{i3}} \bar N^C_{R_i}  N_{R_3} \varphi_1^* 
% \nn\\& 
+\frac{y_{N_{i}}}{\Lambda} \bar N^C_{R_i} N_{R_3} \varphi^*_6 \varphi_5 
+\frac{y_N}{\Lambda} \bar N^C_{R_3} N_{R_3} (\varphi^*_5)^2 
+{\rm c.c.},\\
%%%
V&= \mu_H^2 |H|^2 + \mu_\eta^2 |\eta|^2 + \mu^2_{\varphi_5} |\varphi_5|^2 + \mu^2_{\varphi_6}|\varphi_6|\nn\\
&+
\lambda_H |H|^{4} + \lambda_\eta |\eta|^{4}  + \lambda_{\varphi_5}|\varphi_5|^4  + \lambda_{\varphi_6}|\varphi_6|^4 %+\lambda_{H\eta} (H^\dag H)(\eta^\dag\eta)
+
\lambda_{H\eta} |H|^2 |\eta|^2
+
\lambda'_{H\eta} |H^\dag \eta|^2\nn\\
&+
 \lambda_{H\varphi_5}|H|^2|\varphi_5|^2 + \lambda_{H\varphi_6} |H|^2|\varphi_6|^2
%\nn\\&
+
\lambda_{\eta\varphi_5}|\eta|^2 |\varphi_5|^2+ \lambda_{\eta\varphi_5}|\eta|^2 |\varphi_6|^2
+ \lambda_{\varphi_5\varphi_5}|\varphi_5|^2|\varphi_6|^2
\nn\\&
+ \frac{\lambda''_{H\eta}}{\Lambda} \left((H^\dag\eta)^2 \varphi_6+{\rm h.c.} \right),
\label{eq:lag-lep}
\end{align}
where $\Lambda$ is a cut-off scale, $\tilde H \equiv (i \sigma_2) H^*$ with $\sigma_2$ being the second Pauli matrix, $(a,b)$ runs over $1$ to $3$, and $i$ runs over $1$ to $2$.
Here we take  $\Lambda=$100 TeV in our discussion below, by fixing $g_{BL}(m_{in.}) = g_Y(m_{in.})$. This is just an assumption but it could be reasonable energy scale to discuss low energy scale theory. In detail, see Appendix A.
%%%%%%%%%
\subsection{ Scalar sector}
The scalar fields are parameterized as 
\begin{align}
%\begin{tiny}
&H =\left[\begin{array}{c}
w^+\\
\frac{v + h +i z}{\sqrt2}
\end{array}\right],\quad 
%%%
\eta =\left[\begin{array}{c}
\eta^+\\
\frac{ \eta_R +i \eta_I}{\sqrt2}
\end{array}\right],\quad 
%%%
\varphi_i=
\frac{v_{\varphi_i}+\varphi_{R_i} + iz_{\varphi_i}}{\sqrt2},\ (i=5,6),
\label{component}
%\end{tiny}
\end{align}
where $w^+$ and $z$ are absorbed by the SM gauge bosons $W^+$ and $Z$, and
two massless CP odd bosons $z_{\varphi_5},z_{\varphi_6}$.
%Here we identify the physical goldstone boson (GB) as $z_{\varphi_5}$, therefore $z_{\varphi_6}$ is absorbed by the  $U(1)_{B-L}$ gauge boson $Z_{BL}$.
Then liner combinations of $z_{\varphi_5}$ and $z_{\varphi_6}$ become the physical Goldstone boson (GB) and Nambu-Goldstne boson (NGB) given by
\begin{equation}
\alpha_G = -\sin X z_{\varphi_5} + \cos X z_{\varphi_6}, \quad \alpha_{NG} = \cos X z_{\varphi_5} + \sin X z_{\varphi_6}
\end{equation}
where $\alpha_G (\alpha_{NG})$ is identified as GB (NGB). The mixing angle $X$ is determined from the VEVs of scalar fields:
\begin{equation}
\cos X \equiv \frac{5 v_{\varphi_5}}{\sqrt{25 v_{\varphi}^2 + 36 v_{\varphi_6}^2}}, \quad \sin X \equiv \frac{6 v_{\varphi_6}}{\sqrt{25 v_{\varphi}^2 + 36 v_{\varphi_6}^2}}.
\end{equation} 
%%%%
Inserting tadpole conditions for the CP even matrix in basis of $(\varphi_{R_5},\varphi_{R_6}, h)$,
the mass matrix is given by 
\begin{align}
M_R^2
&\equiv
\left[\begin{array}{ccc}
2 v^2_{\varphi_5} \lambda_{\varphi_5} &  v_{\varphi_5} v_{\varphi_6} \lambda_{\varphi_5\varphi_6} &  v v_{\varphi_5} \lambda_{H\varphi_5} \\ 
v_{\varphi_5} v_{\varphi_6} \lambda_{\varphi_5\varphi_6} & 2 v^2_{\varphi_6} \lambda_{\varphi_6} &  v v_{\varphi_6} \lambda_{H\varphi_6} \\ 
 v v_{\varphi_5} \lambda_{H\varphi_5} & v v_{\varphi_6} \lambda_{H\varphi_6} & 2 v^2 \lambda_{H} \\ 
\end{array}\right],
\end{align}
where we define the mass eigenstate $h_{i}$ ($i=1-3$), and mixing matrix $O_R$ to be $m_{h_{i}}=O_R M_R^2 O_R^T$ and $(\varphi_{R_1},\varphi_{R_2},  h)^T=O_R^T h_i$. Here $h_{SM}\equiv h_3$ is the SM Higgs, therefore,  $m_{h_{SM}}=$125 GeV.
%%%
The mixings among SM Higgs and the other CP-even scalars are constrained by the LHC data 
that suggest their mixing angles should be less than $0.2\sim0.3$ at most~\cite{Chpoi:2013wga, Cheung:2015dta}. Thus,
we assume these mixings are zero to avoid experimental constraints for simplicity,  which can be realized by taking $\lambda_{H \varphi_{5,6}} \ll 1$. 
The mixing between $\varphi_5$ and $\varphi_6$ can be sizable without constraints, but we do not further investigate neutral scalar sector in this work. \\
  %%%%%%%%%%%%%%%%%%%
{\it Inert scalar sector}: Each of the neutral component $\eta_{R/I}$ is given by
\begin{align}
m_{\eta_R}
&= \frac{\mu_\eta^2 +v^2(\lambda_{H\eta}+\lambda'_{H\eta})+ v^2_{\varphi_5} \lambda_{\varphi_5\eta}+ v^2_{\varphi_6} \lambda_{\varphi_6\eta}}{2} + \frac{\lambda''_{H\eta}v^2 v_{\varphi_6}}{2\sqrt2\Lambda},\\
%%%
m_{\eta_I}
&=  \frac{\mu_\eta^2 +v^2(\lambda_{H\eta}+\lambda'_{H\eta})+ v^2_{\varphi_5} \lambda_{\varphi_5\eta}+ v^2_{\varphi_6} \lambda_{\varphi_6\eta}}{2} - \frac{\lambda''_{H\eta}v^2 v_{\varphi_6}}{2\sqrt2\Lambda},
  \end{align}
 where the global minimum at $\langle\eta\rangle=0$ requires the following conditions~\cite{Belanger:2012vp}: 
  \begin{align}
 & 0<(\lambda_H,\ \lambda_\eta),
 \\&
 0<2\sqrt{\frac{\lambda_{H} \lambda_{\eta}}3} + \lambda_{H\eta}+ \lambda'_{H\eta}-|\lambda''_{H\eta}|
\frac{v_{\varphi_6}}{\Lambda},\ 
 %%%
 0<2\sqrt{\frac{\lambda_{\varphi_{5(6)}} \lambda_{\eta}}3} + \lambda_{H\varphi_{5(6)}}.  \end{align}
%%%
Here let us estimate the mass difference between $\eta_R$ and $\eta_I$.  
First of all, let us assume $m_{\eta_R}<m_{\eta_I}=m_{\eta^\pm}$ to evade constraints from the oblique parameters~\cite{Olive:2016xmw}, which implies $\lambda_{H\eta}''\equiv -|\lambda_{H\eta}''|<0$. Then the mass difference can be written by
\begin{align}
\Delta m^2_\eta \equiv m_{\eta_I}^2- m_{\eta_R}^2=\frac{|\lambda_{H\eta}''| v^2 v_{\varphi_6}}{\sqrt2 \Lambda}.
  \end{align}
Once we take $v_{\varphi_6}=1$TeV, we find typical order of  the mass difference such as 
 \begin{align}
\Delta m_\eta \simeq  20.68 \sqrt{|\lambda''_{H \eta}|} \ {\rm GeV},
  \end{align} 
where we have used $v=$246 GeV and $\Lambda=$100 TeV.  If we take $0.01 < |\lambda''_{H \eta}| < 1$ the mass difference is $2.1 \ {\rm GeV} \lesssim \Delta m_\eta \lesssim 21$ GeV.
  
\subsection{$Z'$ boson}  
After $U(1)_{B-L}$ gauge symmetry breaking, we have massive $Z'$ boson. 
The mass is given by
\begin{equation}
m_{Z'} = g_{BL} \sqrt{25 v_{\varphi_5}^2 + 36 v_{\varphi_6}^2},
\end{equation} 
where $g_{BL}$ is the gauge coupling constant for $U(1)_{B-L}$.
Since we take $v_{\varphi_{5,6}}$ to be few TeV the $Z'$ mass becomes $\sim 5$ TeV, taking $g_{BL}$ value as the same as $U(1)_Y$ gauge coupling.
This mass scale is allowed by current LHC search for $Z'$ boson~\cite{Aaboud:2017buh,Klasen:2016qux}, and we omit further analysis.

\subsection{Exotic neutral fermion}
%%%
The masses of neutral fermions are generated by the dimension 5 operators in Eq.~(\ref{eq:Yukawa}). 
The mass matrix for the neutral fermions in basis of $N_{R_{1,2,3}}$ are given by
\begin{align}
M_N=\frac{1}{\sqrt2}
\left[\begin{array}{ccc}
0 & 0  & M_{NS_1}   \\
0 & 0 &M_{NS_2}   \\
M_{NS_1}^*  & M_{NS_2}^*  & M_S   \\
\end{array}\right],
\label{eq-Nmass}
\end{align}
 where we have defined the components as $M_{NS_{1,2}} \equiv y_{N_{1,2}} v_{\varphi_5} v_{\varphi_6}/(\sqrt2 \Lambda)$ and $M_S \equiv y_N v_{\varphi_5}^2/(\sqrt2 \Lambda)$. 
We note that typical order of the components is $\mathcal{O}(1)$ GeV to $\sim 500$ GeV when we take $\Lambda = 100$ TeV, $v_{\varphi_{5,6}}$ as few TeV and $0.1 \lesssim y_{N_{1,2}, N} \lesssim \sqrt{4 \pi}$. 
This matrix is generally diagonalized by 3 by 3 unitary  matrix $V_N$ as $m_{\psi_i}\equiv (V_N M_N V_N^T)_i$ $i=1\sim3$, where $m_{\psi_i}$ is the mass eigenvalue.
Thus the mass eigenstates are given by $N_{R_i} = (V_N^T)_{ij} \psi_{j}$.
The concrete form under the assumption $M_{NS_{i}}^*=M_{NS_{i}}(i=1,2)$  is given by
\begin{align}
m_{\psi}&={\rm diag}\left[0,\frac{\sqrt{4(M_{NS_1}^2+M_{NS_2}^2)+M_S^2}-M_S}2,\frac{\sqrt{4(M_{NS_1}^2+M_{NS_2}^2)+M_S^2}+M_S}2\right],\\
%%%
V_N&=
\left[\begin{array}{ccc}
1 & 0  & 0   \\
0 & i &0   \\
0  & 0 & i   \\
\end{array}\right]
%%%
\left[\begin{array}{ccc}
-\frac{M_{NS_2}}{\sqrt{m_{\psi_2}m_{\psi_3}}} & -\frac{M_{NS_1}}{\sqrt{m_{\psi_2}(m_{\psi_2}+m_{\psi_3})}}  & \frac{M_{NS_1}}{\sqrt{m_{\psi_3}(m_{\psi_2}+m_{\psi_3})}}    \\
%%%
\frac{M_{NS_1}}{\sqrt{m_{\psi_2}m_{\psi_3}}} & -\frac{M_{NS_2}}{\sqrt{m_{\psi_2}(m_{\psi_2}+m_{\psi_3})}}  & \frac{M_{NS_2}}{\sqrt{m_{\psi_3}(m_{\psi_2}+m_{\psi_3})}}    \\
%%%
0  & \sqrt{\frac{m_{\psi_2}}{m_{\psi_2}+m_{\psi_3}}} & \sqrt{\frac{m_{\psi_3}}{m_{\psi_2}+m_{\psi_3}}}   \\
\end{array}\right],
\end{align}
where $V_N$ is orthogonal matrix under $M_{NS}^*=M_{NS}$.

%%%%%%%%%%%%%%%%%%%
\begin{figure}[t]
\begin{center}
%\hspace{3cm}
\includegraphics[width=100mm]{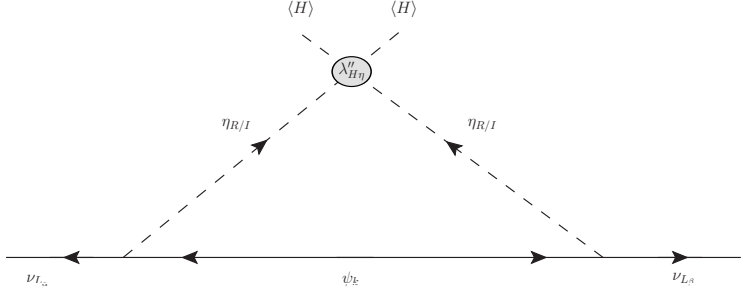} 
\end{center}
\caption{The one loop diagram which induces neutrino masses. } 
  \label{fig:diagram}
\end{figure}
%%%%%%%%%%%%%%%%%%%

  \subsection{Masses for the lepton sector}
The charged lepton masses are given by $m_\ell =y_\ell v/\sqrt2$ after the electroweak symmetry breaking, where $m_\ell$ is assumed to be the mass eigenstate.
 %%%%%%%%%
The neutrino mass matrix is induced at the one-loop level in Fig.~\ref{fig:diagram}~\cite{Ma:2006km}, and its formula is given by
  \begin{align}
&({\cal M}_{\nu})_{\alpha\beta}  =
\sum_{k=2}^3 \frac{Y_{\alpha k} m_{\psi_{k}} (Y^T)_{k\beta}} {(4\pi)^2}
\left[
\frac{m^2_{\eta_R}}{m^2_{\eta_R}-m^2_{\psi_{k}}}\ln\frac{m^2_{\eta_R}}{m^2_{\psi_{k}}}
-
\frac{m^2_{\eta_I}}{m^2_{\eta_I}-m^2_{\psi_{k}}}\ln\frac{m^2_{\eta_I}}{m^2_{\psi_{k}}}
\right],
  \end{align}
where we redefine $Y_{a \alpha}\equiv \sum_{i=2}^3\frac{y_{\nu_{ai}} V^T_{N_{i\alpha}}}{\sqrt2}$.
\footnote{There is a model that $\lambda''_{H\eta}$ is generated at one-loop level~\cite{Aoki:2017eqn, Fukuoka:2009cu}. }
{\it Notice here that massless fermion $\psi_1$ does not contribute to the neutrino masses and  their oscillations. Thus components $Y_{a1}$ do not get any constraints from neutrino mass.}
%and $y_\nu'$ is three by three Yukawa matrix extension of $y_\nu$  by inserted null column vector for the third column.
  %%%
 Once we define $D_\nu\equiv U_{MNS} {\cal M}_{\nu} U_{MNS}^T \equiv U_{MNS} (Y R Y^T)  U_{MNS}^T$,
 $(Y)_{3\times2}$ can be rewritten in terms of observables and several arbitral parameters as:
 \begin{align}
 Y_{\alpha i} = \left( U^\dag_{MNS} D_\nu^{1/2} O R^{-1/2} \right)_{\alpha i},\quad 
 R\equiv \frac{m_{\psi}}{(4\pi)^2 } 
 %%%
 \left[
\frac{m^2_{\eta_R}}{m^2_{\eta_R}-m^2_{\psi}}\ln\frac{m^2_{\eta_R}}{m^2_{\psi}}
-
\frac{m^2_{\eta_I}}{m^2_{\eta_I}-m^2_{\psi}}\ln\frac{m^2_{\eta_I}}{m^2_{\psi}}
\right],\label{eq:neut-rel}
  \end{align}
where $O\equiv O(z)$, satisfying $O^TO=1_{2\times2}$ but $OO^T=$diag(0,1,1), is an arbitral 3 by 2 matrix with complex value of $z$, and $U_{MNS}$ and $D_\nu$ are obsrevables~\cite{Gonzalez-Garcia:2014bfa}.
%%%
%From Eq.~(\ref{eq:neut-rel}), the concrete structure of $Y_{\nu}$ can be found to be
Depending on the mass ordering of active neutrinos, $O$ can concretely be parametrized by
\begin{align}
O =
\left[\begin{array}{cc}
 0  & 0   \\
 \cos z & -\sin z   \\
 \pm \sin z &  \pm \cos z    \\
\end{array}\right],\ {\rm for\  NH},\
%%%
\left[\begin{array}{cc}
 \cos z & -\sin z   \\
 \pm \sin z &  \pm \cos z    \\
  0  & 0   \\
\end{array}\right],\ {\rm for\ IH},\
\label{eq:ykw}
\end{align} 
where NH(IH) is short-hand notation of "Normal(Inverted) Hierarchy", and the lightest active neutrino mass is zero. 
%where only the components $\times$ are nonzero, its form is independent of active neutrino hierarchy.
%This is one of the important consequences as we will see below.

\subsection{Lepton flavor violations} 
LFV processes $\ell \to \ell' \gamma$ are induced from the neutrino Yukawa couplings at one-loop level, and their forms are given by
\begin{align}
BR(\ell_\alpha\to \ell_\beta \gamma)&\approx\frac{4\pi^3\alpha_{em}C_{\alpha\beta}}{3(4\pi)^4G_F^2}
\left|\sum_{i=2}^3(Y)_{\beta i} (Y^\dag)_{i\alpha} F_{lfv}(\psi_{i},\eta^\pm)\right|^2,\\
%%%
F_{lfv}(a,b)&\equiv\frac{2 m_a^6+3m_a^4m_b^2-6m_a^2m_b^4+m_b^6+12m_a^4m_b^2\ln\left[\frac{m_b}{m_a}\right]}{(m_a^2-m_b^2)^4},
\end{align}
where $\alpha_{em}\approx1/134$ is the fine-structure constant, $G_F\approx1.17\times10^{-5}$ GeV$^{-2}$ is the Fermi constant,
and $C_{21}\approx1$, $C_{31}\approx 0.1784$, $C_{32}\approx0.1736$. 
The stringent constraint comes from $\mu\to e \gamma$ and its upper bound is given by $BR(\mu\to e\gamma)\lesssim 4.2\times10^{-13}$~\cite{TheMEG:2016wtm}.
%%%
%Reminding the structure of $Y_\nu$ in Eq.~(\ref{eq:ykw}), one finds ${\rm BR}(\mu\to e \gamma) ={\rm BR}(\tau\to e \gamma)=0$.
%Therefore we consider the contribution of ${\rm BR}(\tau\to \mu \gamma)$ only and its experimental upper bound is given by~\cite{ Hayasaka:2007vc}:  ${\rm BR}(\tau\to \mu \gamma)\lesssim 4.4\times 10^{-13}$, where we define $\ell_1\equiv e$,  $\ell_2\equiv \mu$, and  $\ell_3\equiv \tau$.

\subsection{ Dark matter}
 Here we discuss if our model can have viable DM candidate. 
In general, this class of model has two kinds of DM candidates; the lightest fermion $m_{\psi_1}$ and/or the lightest neutral inert boson $m_{\eta_{R}}$.
However since the lightest fermion is massless, an inert boson is in favor of being  the good DM candidate.
Even when it is the case, one might worry about the too fast decay of DM;
the decay mode $\eta_R\to \nu_L \psi_1$ arises from $y_\nu$, and its decay rate is written by
$\Gamma\sim \frac{m_{\eta_R}}{16\pi^2} \sum_{a=1}^3 |(Y)_{a1}|^2$. 
%%%
Then we evaluate the upper bound on $\Gamma$ by imposing its lifetime $\tau$ ($\Gamma=\tau^{-1}$) should be longer than the current Universe,
therefore we have the following constraint:
\begin{align} 
\Gamma \lesssim 1.51\times 10^{-42}\ {\rm GeV}.
\end{align}
Although we can take $ \sum_{a=1}^3 |(Y)_{a1}|^2$ as free parameters because it does not contribute to the neutrino oscillation data, the above constraint severely restricts our model. 
 In addition, we need to take into account quantum corrections to the couplings to check the constraint can be satisfied at low scale.
In principle, our requirement can be realized by tuning the free parameters.
Therefore one can have viable DM candidate if we imposing fine tuning for the Yukawa couplings.

If one assumes that there exists DM in our theory, which is inert boson $\eta_R$,
one has to rely on modes via kinetic term and/or Higgs potential to explain the relic density of DM; $\Omega h^2\approx$0.12~\cite{Ade:2013zuv}.
When the mode from Higgs portal interaction is subdominant due to constraint from DM direct detection searches, the main mode comes from gauge interactions in kinetic term. 
 Note that we have $Z'$ interaction of inert boson $\eta_R$ as it is charged under the extra $U(1)$. However the $Z'$ interaction will be subdominant since we should require heavy $Z'$ mass as $\gtrsim$ few TeV or small gauge coupling $g_{BL}$ due to current LHC constraints. In addition, the  $Z'$ interaction takes the form of $Z'_\mu (\partial^\mu \eta_R \eta_I - \eta_R \partial^\mu \eta_I)$ and it does not contribute to DM-nucleon scattering if masses of $\eta_R$ and $\eta_I$ are not much degenerated
 ; $Z'$ exchange contributes to DM-nucleon scattering if $\eta_R$ and $\eta_I$ have degenerate masses~\footnote{The mass difference should be larger than the order 100 keV originated from the typical kinetic energy of DM around the earth. If not, a strong bound has to be imposed from direct detection experiments. See for example ref.~\cite{Ma:2015mjd} for the case where DM is complex scalar whose real and imaginary part have the same mass.} .  
Hence  the DM feature is almost the same as two Higgs doublet model with an inert Higgs, which has already been discussed in ref.~\cite{Hambye:2009pw}. Therefore the allowed mass is at around $\gtrsim 500$ GeV, once the W/Z final state modes are opened.

\subsection{The effective number of neutrino species: $\Delta N_{\rm eff}$} 
The massless fields contribute to the relativistic energy density of Universe, which is denoted by $\Delta N_{\rm eff}$.
A thermalized scalar(fermion) contributes $\Delta N_{\rm eff}=4/7(1)$, each of which is consistent with bounds from Big Bang Nucleosynthesis (BBN),
which are in the range of $\Delta N_{\rm eff}<1.2$~\cite{Mangano:2011ar} at 95\% CL, depending on the primordial abundances. 
%%%
Moreover, once we assume that these fields typically decouple from the plasma at temperatures above the QCD phase transition $\sim\cal O$(100) MeV,
we find the effective number of relativistic degrees of freedom to be about 60.
Therefore we obtain
\begin{align}
\Delta N_{\rm eff}\lesssim \frac{11}{7} \left(\frac{10.75}{60}\right)^{4/3}\approx0.159.
 \end{align}
 This value is still in good agreement with 
the recent experimental data such as Planck $\Delta N_{\rm eff}=0.15\pm0.23$~\cite{Ade:2015xua}.
Note that although massless particles are charged under $B-L$ they interact with the SM particles by exchanging heavy $Z'$ and/or scalar bosons which are $\mathcal{O}(100)$ GeV or $\mathcal{O}(1)$ TeV scale. 
Thus our massless particles can be decoupled at the early Universe before QCD phase transition. Here BBN occurs after QCD phase transition.

\subsection{Numerical analysis}
%------------------------------------------------------------------------
\begin{figure}[t]
\centering
\includegraphics[width=7cm]{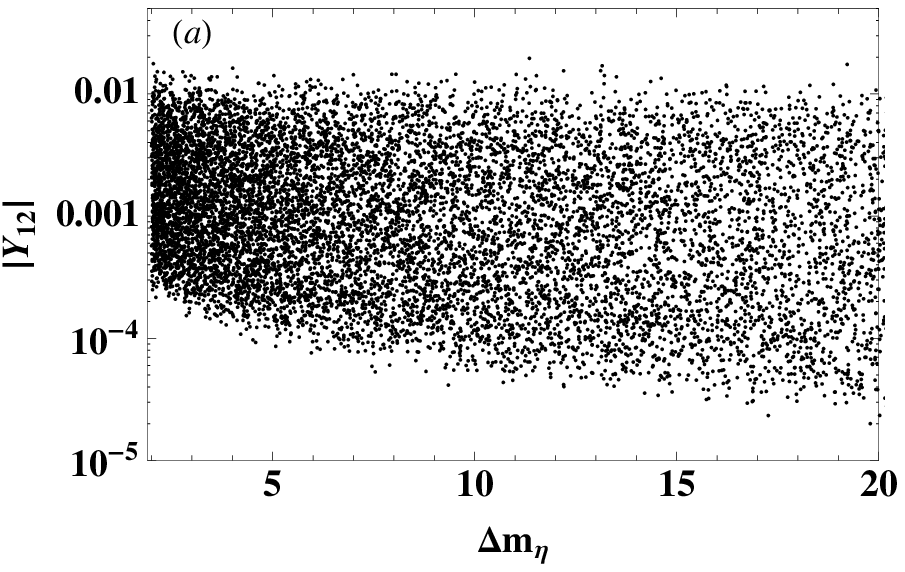}
\includegraphics[width=7cm]{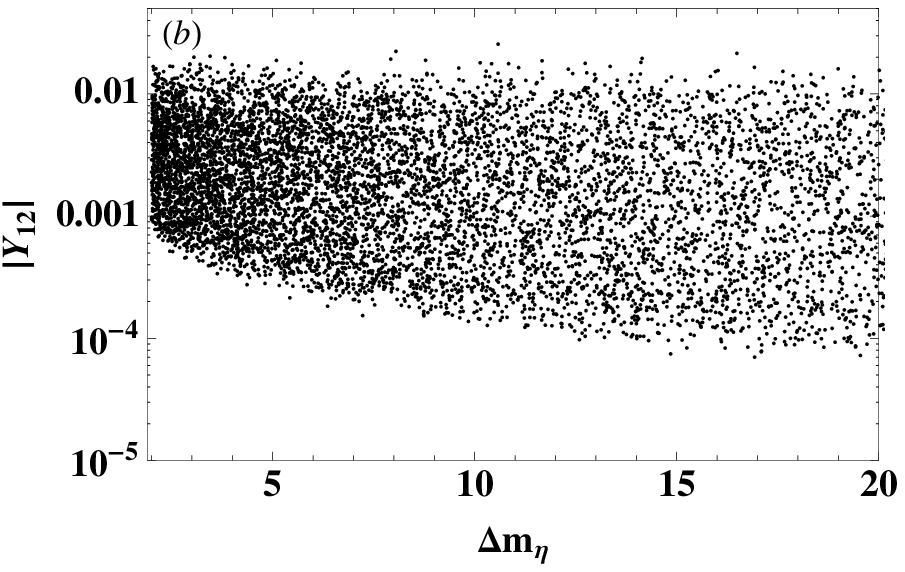}
\includegraphics[width=7cm]{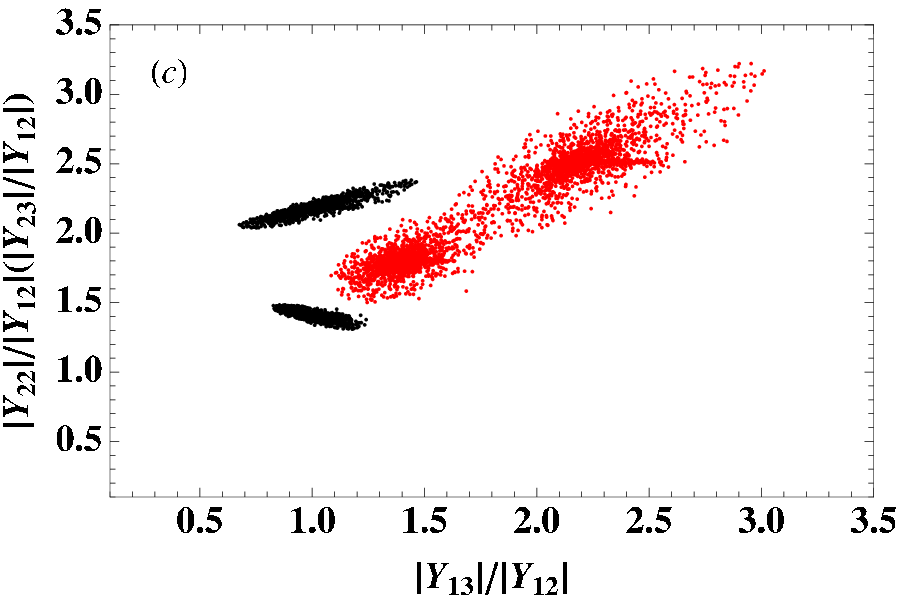}
\includegraphics[width=7cm]{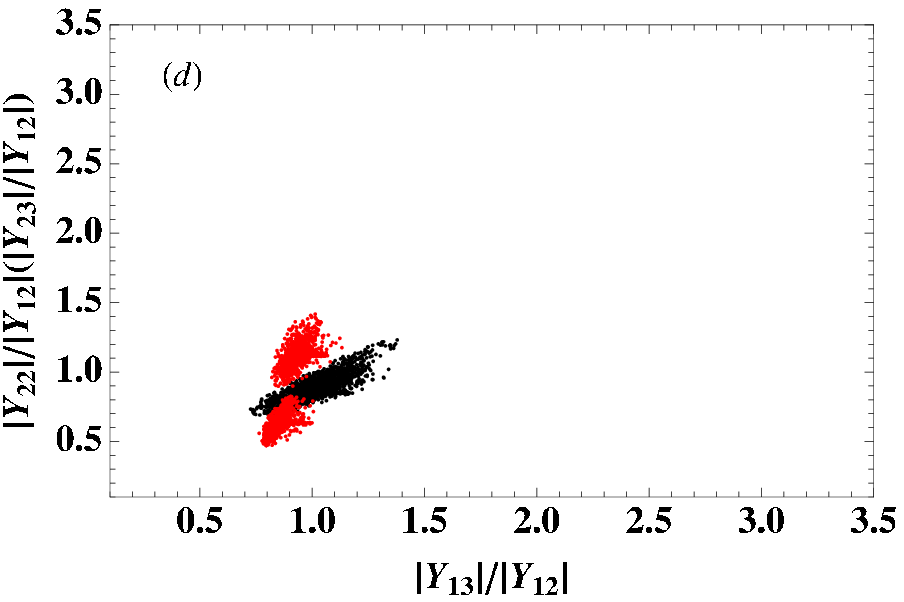}
\includegraphics[width=7cm]{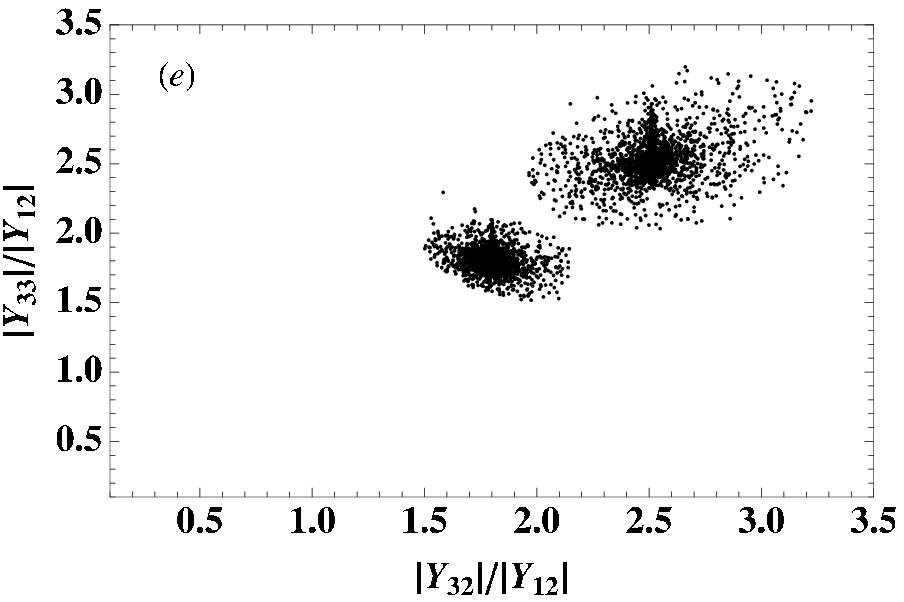}
\includegraphics[width=7cm]{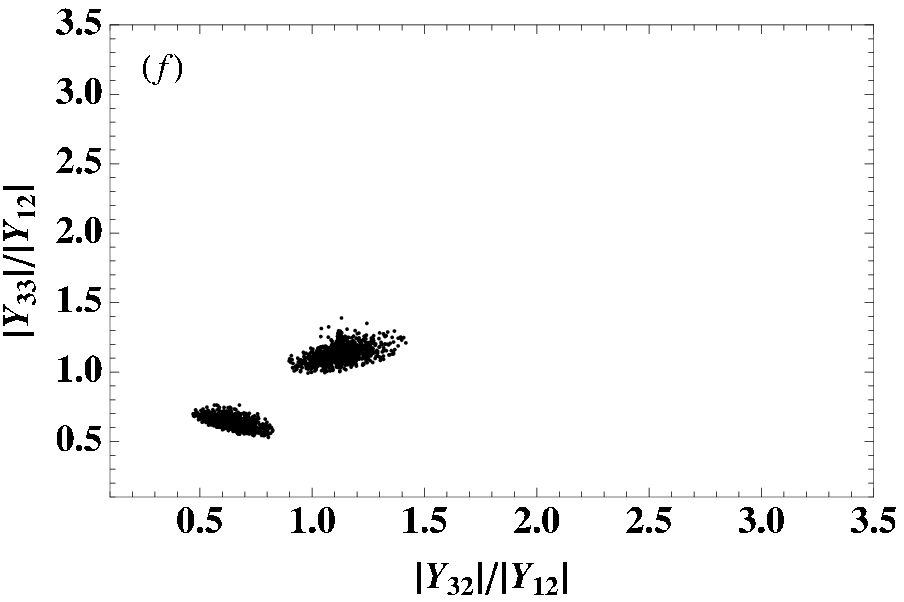}
\caption{(a) and (b): Correlation between $Y_{12}$ and $\Delta m_\eta$ for NH (a) and IH (b) cases. (c) and (d): correlation between $|Y_{13}|/|Y_{12}|$ and $|Y_{22(23)}|/|Y_{12}|$ by balck(red) dots for NH (c) and IH (d) cases. (e) and (f): correlation between $|Y_{32}|/|Y_{12}|$ and $|Y_{33}|/|Y_{12}|$ for NH (e) and IH (f) cases.}
\label{fig:Yukawa}
\end{figure}
\begin{figure}[t]
\centering
\includegraphics[width=7cm]{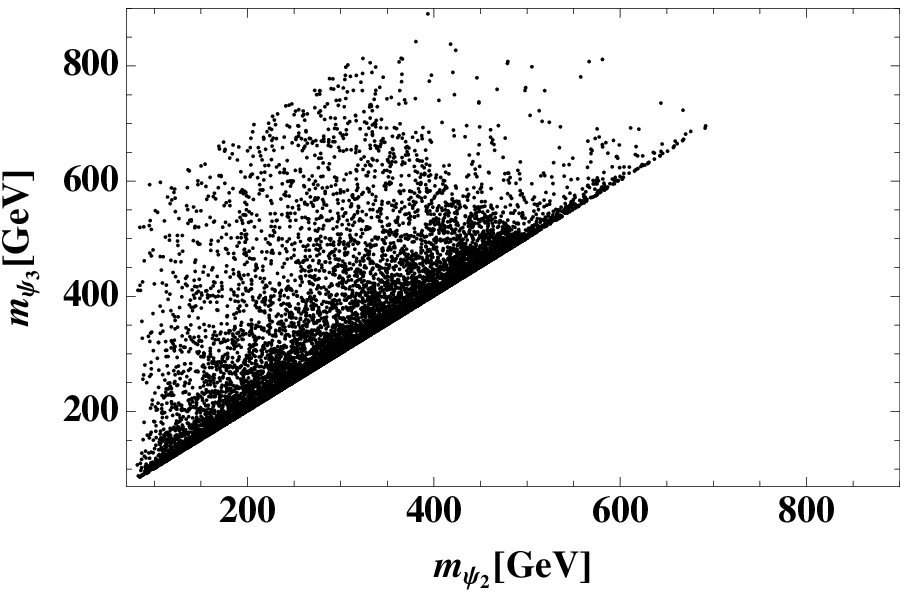}
\includegraphics[width=7cm]{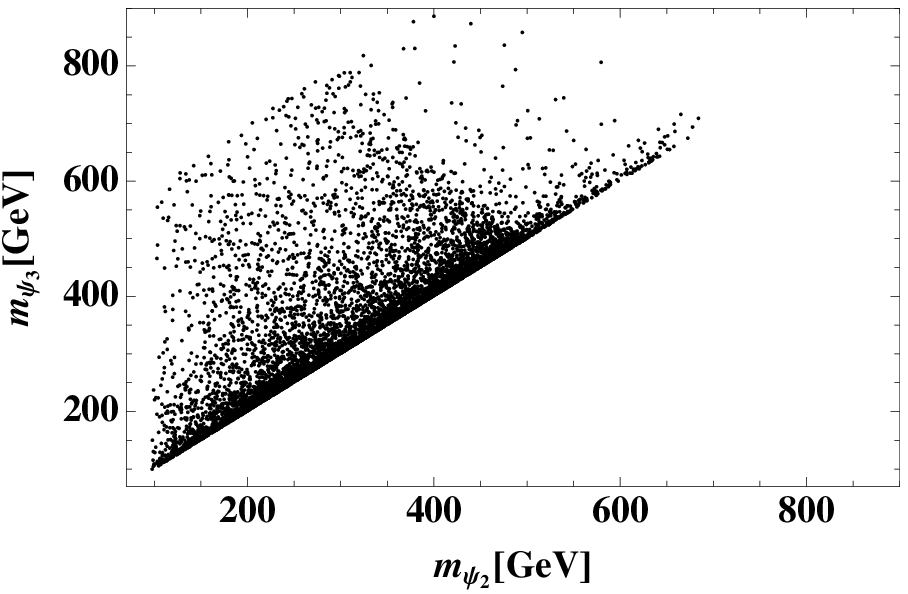}
\caption{Scattering plots to satisfy the neutrino oscillation data and LFVs in terms of the correlation between $m_{\psi_1}$ and $m_{\psi_2}$,
where the the left(right) figure represents the case of NH(IH).
}
\label{fig:m2-m3}
\end{figure}
\begin{figure}[t]
\centering
\includegraphics[width=7cm]{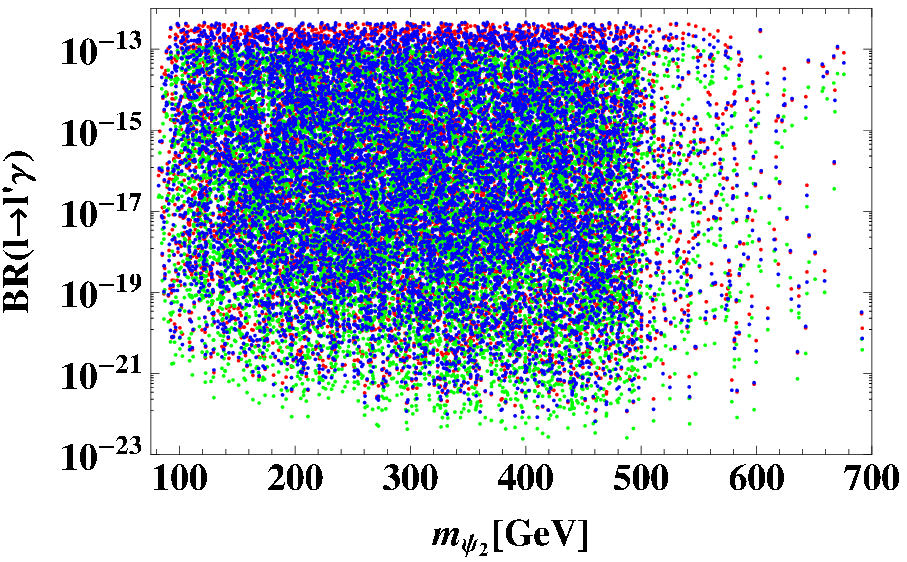}
\includegraphics[width=7cm]{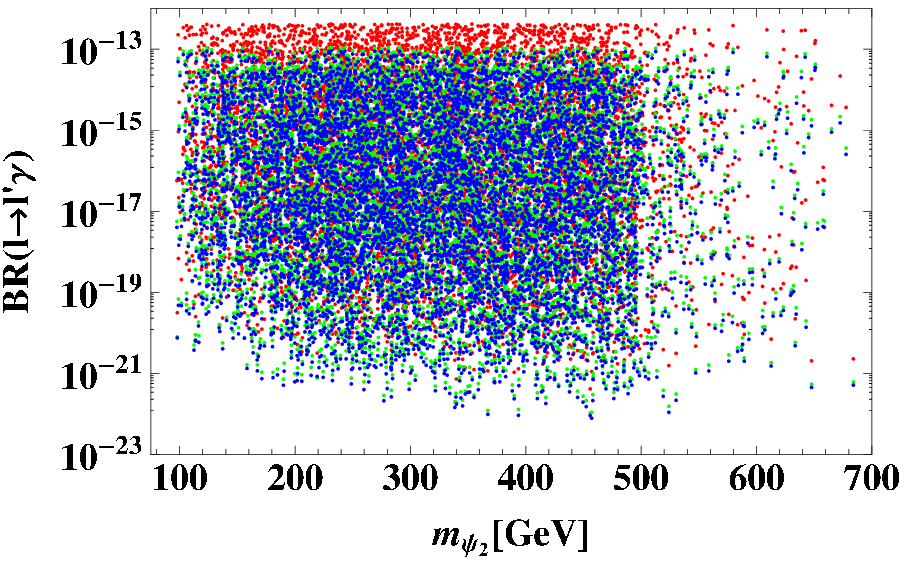}
\caption{Scattering plots in terms of the correlation between the values of LFVs and $m_{\psi_2}$,
where the the left(right) figure represents the case of NH(IH). The points of  red, green, and blue respectively represent the case of $BR(\mu\to e\gamma)$, $BR(\tau\to e\gamma)$, and $BR(\tau\to \mu\gamma)$.
}
\label{fig:m2-lfvs}
\end{figure}
\begin{figure}[t]
\centering
\includegraphics[width=7cm]{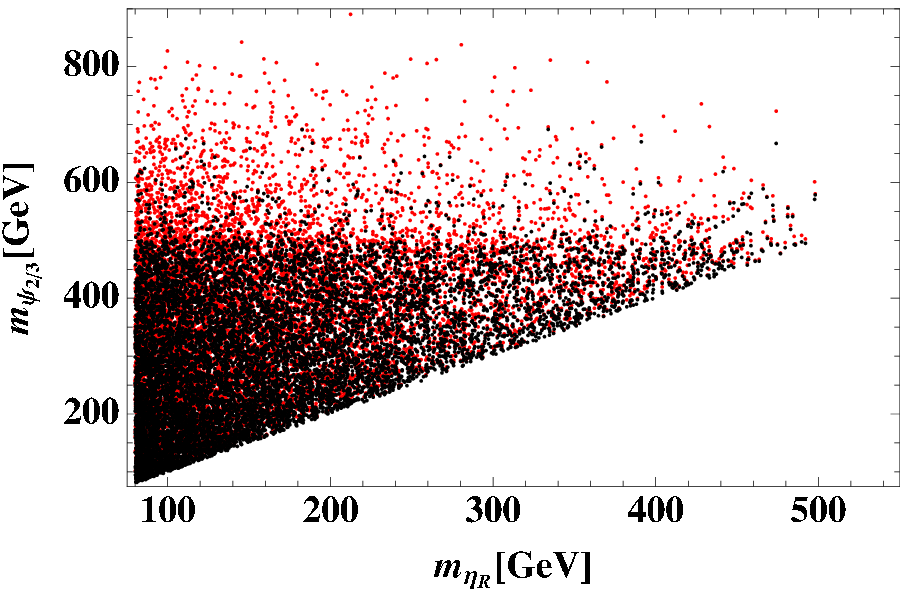}
\includegraphics[width=7cm]{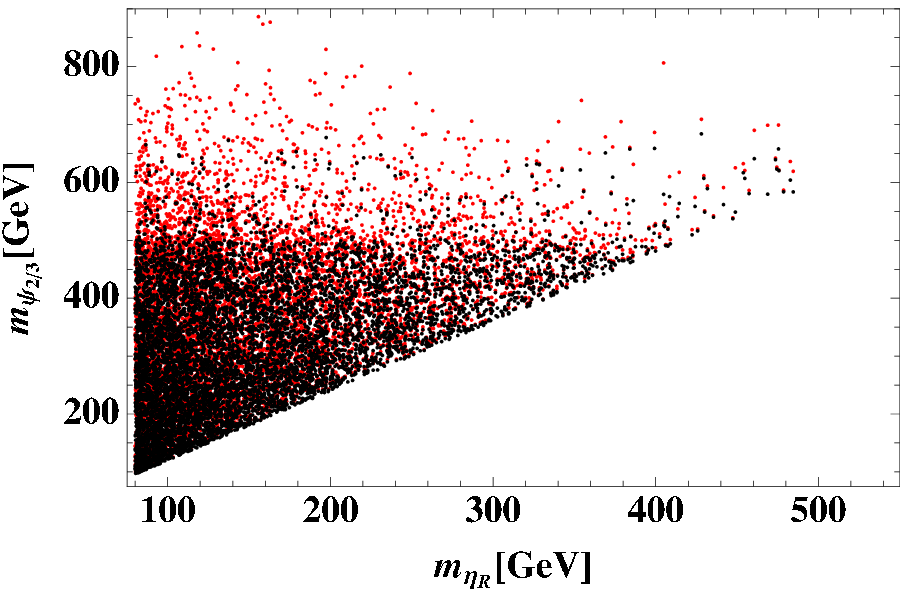}
\caption{Scattering plots to satisfy the neutrino oscillation data and LFVs in terms of the correlation between $m_{\eta_R}$ and $m_{\psi_{2/3}}$,
where the the left(right) figure represents the case of NH(IH). The points of  black and red respectively represent the case of $m_{\psi_2}$ and $m_{\psi_3}$.
}
\label{fig:mx-m23}
\end{figure}

%------------------------------------------------------------------------
Here we explore the allowed region to satisfy neutrino oscillation data and constraint from BR$(\ell \to \ell' \gamma)$.
 First of all, we randomly select the following input parameters:
\begin{align}
& {\rm Re}[z]\in (0,\pi),\quad {\rm Im}[z]\in (1,10), \nonumber \\
& (M_{NS_{1,2}}, M_{S}) \in (1,500) \ {\rm GeV}, \quad m_{\eta_R} \in (100, 500) \ {\rm GeV}, \nonumber \\
& \Delta m_\eta \in (2, 20) \ {\rm GeV}, \quad m_{\eta^\pm} = m_{\eta_I} = \sqrt{m_{\eta_R}^2 + \Delta m_\eta^2},
\end{align}
where we take the typical region of the parameters as discussed above. 
Here we also apply a condition $m_{\eta_R} < m_{\psi_{2,3}}$ expecting $\eta_R$ to be a DM candidate.
In addition we apply best fit values of the current neutrino oscillation data for NH and IH cases~\cite{Olive:2016xmw}. As our outputs we obtain $Y_{\alpha i}$ ($\alpha=1-3$, $i=1,2$) from our formula Eq.~(\ref{eq:neut-rel}).
%~\footnote{we have analyzed the case of inverted hierarchy, but the result does not change much.} 
 Firstly, we show correlation between the size of Yukawa coupling $|Y_{12}|$ and $\Delta m_\eta$ in Fig.~\ref{fig:Yukawa}-(a) and -(b) for NH and IH cases. Moreover relative size of the Yukawa couplings $|Y_{\alpha i}|/|Y_{12}|$ are shown in Fig.~\ref{fig:Yukawa}-(c,e) and -(d,f) for NH and IH cases. From the figures we see the typical size of $|Y_{\alpha i}|$ is $\mathcal{O}(10^{-4})$ to $\mathcal{O}(10^{-2})$. Thus original couplings in the Lagrangian $y_\nu$ have the same order of values and it is similar to the size of $y_\ell$ which provide charged lepton masses. Also we find that the correlations among the Yukawa couplings are clearly different between NH and IH cases.
Fig.~\ref{fig:m2-m3} shows  the scattering plots to satisfy the neutrino oscillation data and LFVs in terms of the correlation between $m_{\psi_1}$ and $m_{\psi_2}$, where the the left(right) figure represents the case of NH(IH).
The Fig.~\ref{fig:m2-lfvs} shows  the scattering plots in terms of the correlation between the values of LFVs and $m_{\psi_2}$,
where the left(right) figure represents the case of NH(IH). The points of  red, green, and blue respectively represent the case of $BR(\mu\to e\gamma)$, $BR(\tau\to e\gamma)$, and $BR(\tau\to \mu\gamma)$.
The Fig.~\ref{fig:mx-m23} shows  the scattering plots to satisfy the neutrino oscillation data and LFVs in terms of the correlation between $m_{\eta_R}$ and $m_{\psi_{2/3}}$,
where the left(right) figure represents the case of NH(IH). The points of  black and red respectively represent the case of $m_{\psi_2}$ and $m_{\psi_3}$.
These figures suggest that there are no difference between NH and IH for the masses of $m_{\psi_{2,3,\eta_R}}$,
while the LFVs give differences among each processes;  upper bound on $\tau\to e\gamma$ in the NH case is lower than the other two processes,
while upper bound on $\mu\to e\gamma$ in the IH case is higher than the other two processes by half.
And orders of upper bounds for three processes  for NH and IH are respectively found to be $10^{-13}$ and $10^{-14}$.
%The left side of Fig.~\ref{fig:neut-lfvs} shows the correlation between $m_{\psi_1}$ and $m_{\psi_2}$, while the The right side of Fig.~\ref{fig:neut-lfvs} shows the correlation between $m_{\eta_R}$ and BR$(\tau\to\mu e)$. These figures show that there are still wide ranges of mass parameters, even though the Yukawa couplings are restricted by our model. 
%{\it Notice here that the allowed range of $m_{\eta_R}$ includes the narrow region of DM mass~$\sim$500 GeV.}

%------------------------------------------------------------------------
%------------------------------------------------------------------------

%------------------------------------------------------------------------
%------------------------------------------------------------------------

%------------------------------------------------------------------------
%------------------------------------------------------------------------

\section{Conclusion}
We have proposed a model based on an alternative $U(1)_{B-L}$ gauge symmetry with 5 dimensional operators in the Lagrangian in which we have constructed the neutrino masses at one-loop level introducing minimal field contents, and discussed LFVs, DM, and $\Delta N_{\rm eff}$. 
Then numerical analysis is carried out to search for values of parameters accommodating observed data adopting some input parameter without tuning.
As a result we have found allowed region to satisfy all the data such as neutrino oscillation data without conflict of several constraints such as LFV and $\Delta N_{\rm eff}$.
%which we have discussed above.
Below we list several remarks:
%%%
\begin{enumerate}
\item 
To estimate our cut-off scale, we have assumed that the initial value of gauge coupling $g_{BL}$ at $m_Z$ is same as the one of hypercharge $g_Y$; $g_{BL}(m_Z)=g_Y(m_Z)$. Then we have obtained $\Lambda=100$ TeV, then the mass difference between $\eta_R$ and $\eta_I$
is of the order 0.1 GeV under $v=246$ GeV, $v_{\varphi_6}=1$ TeV, and $\lambda''_{H\eta}=1$.
Since it contributes to the neutrino masses, more natural parametrization has been achieved to explain the scale of neutrino masses.
Of course, we can always increase the value of $\Lambda$, by enlarging the initial value of $g_{BL}$.
%%%

\item
Lightest active neutrino is massless that arises from rank two matrix of $\psi$.
%%%
Through our numerical analyses, there are no difference between NH and IH for the masses of $m_{\psi_{2,3,\eta_R}}$ 
 while we have found correlations among relative sizes of the Yukawa couplings related to $\bar L_L \tilde \eta N_R$ terms. Also typical order of the Yukawa couplings is $\mathcal{O}(10^{-4})$ to $\mathcal{O}(10^{-2})$ which is similar to SM Yukawa couplings for charged lepton masses.  In addition 
the LFVs give differences among each processes;  upper bound on $\tau\to e\gamma$ in the NH case is lower than the other two processes,
while upper bound on $\mu\to e\gamma$ in the IH case is higher than the the other two processes by half.
And orders of upper bounds for three processes  for NH and IH are respectively found to be $10^{-13}$ and $10^{-14}$.

\item
 DM candidate in the model is neutral component of inert scalar doublet since the lightest component of $\psi$ is massless. 
However inert boson can decay into $\psi_1 \nu_L$ via Yukawa interaction. 
We find that one can in principle stabilize the DM candidate by fine-tuning the parameters since the Yukawa interactions related to DM are free parameters due to absence of neutrinos masses.
In addition relic density of DM can be realized as in the inert Higgs doublet model.
Thus it is possible to accommodate DM in our model if we require fine-tuning for the Yukawa couplings.

%%%
\item 
We have simply estimated  $\Delta N_{\rm eff}$, since we have a physical massless fermion in addition to the massless boson.
 And we have confirmed it is still within experiment by Planck.
\end{enumerate}

Note that relative size of Yukawa couplings $Y_{\alpha i}$ is related to decay branching ratio of $\eta^\pm$ which decays into $\ell^\pm \Psi_i$ via Yukawa interactions.
Thus prediction to the correlation among the Yukawa couplings would be tested by searching for signals from $\eta^\pm$ production at collider like LHC. 
%%%
%{\color{red}Before closing the section, it is worthwhile to mention the possibility of ultraviolet completion, which can be achieved by providing a concrete structure of $\Lambda$. To realize such a scenario, one might introduce additional symmetries with new fields that mediate inside the loop of $\Lambda$. For example of the successful model, see ref.~\cite{Aoki:2013gzs}.}

\appendix

\section{ Beta function of $g_{BL}$}
\label{beta-func}
\begin{figure}[t]
\centering
\includegraphics[width=10cm]{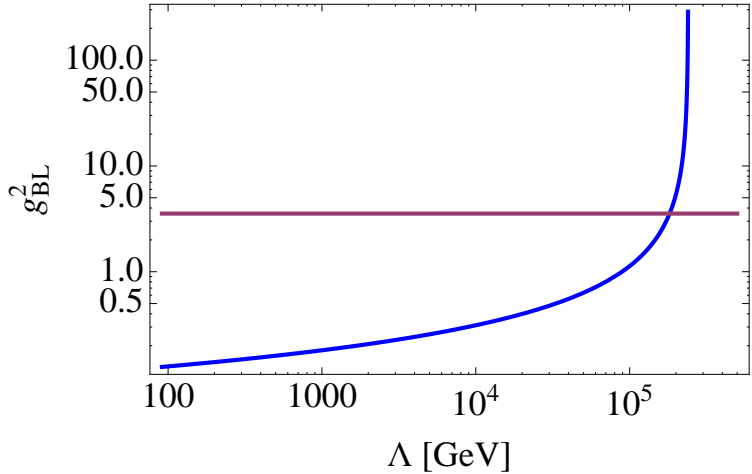}
\caption{RGE flow of $g_{BL}$ in terms of cut-off scale $\Lambda$, where $g_Y(m_{in.})=g_{BL}(m_{in.})$ is assumed. }
\label{fig:rge}
\end{figure}

Here we discuss running of $g_{BL}$ coupling and estimate the effective energy scale by evaluating the Landau pole due the presence of new fields.
Each of $B-L$ beta function for boson and fermion is given by
\begin{align}
 b^{b}_{B-L}=33 ,\quad
%%%
 b^{f}_{B-L}=\frac{140}{3}.
\end{align}
%%%
Then one finds the following energy evolution of the gauge coupling:
\begin{align}
\frac{1}{g^2_{B-L}(\Lambda)}&=\frac1{g_{B-L}^2(m_{in.})}
-\theta(\Lambda-m_{th.}) \frac{ b^b_{B-L}+ b^f_{B-L}}{(4\pi)^2}\ln\left[\frac{\Lambda^2}{m_{th.}^2}\right],\label{eq:rge_gy}
\end{align}
where $\Lambda$ is a reference energy, and we assume to be $m_{in.}(=m_Z)<m_{th.}=$100 GeV, with the same threshold masses $m_{th.}$ for fermions and bosons.
%%%
%Here we include contributions from three exotic singlet fermions $E$, new doublet scalar $\Phi_{N/2}(\Phi'_{N'/2})$, and a doubly charged gauge singlet scalar $k^{\pm\pm}$ that plays a crucial role in making appropriate decays into the SM fields as we will discuss later.
Once we fix to be $g_{BL}(m_{in.}) = g_Y(m_{in.})$, we obtain the RGE flow as can be seen in Fig.~\ref{fig:rge}.
%%%
It shows that $g_{BL}$ is valid up to around 100 TeV.
Notice here that RGE is very sensitive to the initial value of $g_{BL}$, and we can always enlarge the cut-off scale by decreasing the value of $g_{BL}$.
For example if one fix to be $g_{BL}(m_{in.})=0.1\times g_Y(m_{in.})$, then the theory is valid up to Plank mass scale $\sim 10^{18}$ GeV.

%------------------------------------------------------------------------
%------------------------------------------------------------------------

%\section{ Physical Goldstone boson and Nambu Goldstone boson}

\end{document}